# Like-for-like bibliometric substitutes for peer review: advantages and limits of indicators calculated from the $e_p$ index

Alonso Rodríguez Navarro[a,b]*, Ricardo Brito[b]

[a] *Departamento de Biotecnología-Biología Vegetal, Universidad Politécnica de Madrid, Avenida Puerta de Hierro 2, 28040, Madrid, Spain*
[b] *Departamento de Estructura de la Materia, Física Térmica y Electrónica and GISC, Universidad Complutense de Madrid, Plaza de las Ciencias 3, 28040, Madrid, Spain*

*\* Corresponding author e-mail address: alonso.rodriguez@upm.es*

The use of bibliometric indicators would simplify research assessments. The 2014 Research Excellence Framework (REF) is a peer review assessment of UK universities, whose results can be taken as benchmarks for bibliometric indicators. In this study we use the REF results to investigate whether the $e_p$ index and a top percentile of most cited papers could substitute for peer review. The probability that a random university's paper reaches a certain top percentile in the global distribution of papers is a power of the $e_p$ index, which can be calculated from the citation-based distribution of university's papers in global top percentiles. Making use of the $e_p$ index in each university and research area, we calculated the ratios between the percentage of 4-star-rated outputs in REF and the percentages of papers in global top percentiles. Then, we fixed the assessment percentile so that the mean ratio between these two indicators across universities is 1.0. This method was applied to four units of assessment in REF: Chemistry, Economics & Econometrics joined to Business & Management Studies, and Physics. Some relevant deviations from the 1.0 ratio could be explained by the evaluation procedure in REF or by the characteristics of the research field; other deviations need specific studies by experts in the research area. The present results indicate that in many research areas the substitution of a top percentile indicator for peer review is possible. However, this substitution cannot be made straightforwardly; more research is needed to establish the conditions of the bibliometric assessment.

# 1. Introduction

Research investments in technologically advanced countries are quite high and research policy must control these investments by both boosting the lines of research that have the most economic or societal importance (Weinberg 1962, 1964) and checking the returns to society (Salter and Martin 2001), which includes the assessing of the research performance. Although performance assessments in research are not more necessary than in any other productive system, in research the procedure is more complex: "A factory can easily measure how many widgets are produced per man-hour of labor. Evaluating scientific productivity, however is trickier" (Kreiman and Maunshell 2011, p. 1). The consequence is that wrong research evaluations have been frequent, giving rise to notable mistakes, as in the well-known case of the European paradox (Bonaccorsi 2007; Dosi et al. 2006; Herranz and Ruiz-Castillo 2013; Rodriguez-Navarro and Narin 2018). The conceptual problem that explains these mistakes lies in the fact that the product of a research system, the advancement of knowledge, is an intangible product that cannot be easily measured. From a rational point of view, the best judges to evaluate the intangible advancement of science are the same researchers that produce it. But an assessment in which the same actors are judge and party does not seem to be the best solution. Judges that are sufficiently expert as to perform a competent assessment and sufficiently distant as to avoid conflicts of interest can be selected, but to organize a research assessment by this method is complex and onerous (Martin 2011; Régibeau and Rockett 2016).

To evaluate the performance of a research system indirectly, without actually assessing its contribution to the advancement of knowledge, a whole field of science has been developed: *scientometrics*, a branch of which, *bibliometrics*, uses numerical analyses of scientific publications and their citations. This field of science has developed numerous indicators that can be used as proxies of the advancement of knowledge (De-Bellis 2009; Godin 2006; Mingers and Leydesdorff 2015; Waltman 2016). However, because the number of publications and citations is large and these data can be easily obtained in several databases, it is easy to produce indicators using intuition, imagination, or mathematical skills, but which are not necessarily indicators of scientific progress.

The difficulty of producing reliable indicators of scientific progress arises from the fact that a large proportion of the scientific publications are "normal science"; these publications are necessary for the progress of knowledge but are not part of it (Rodríguez-Navarro 2012). Namely, they give rise to but are not part of the scarce "revolutionary science" that boosts knowledge (Kuhn 1970); it has been calculated that less than 0.02% of all publications are landmark publications (Bornmann et al. 2018). These landmark publications cannot be counted in most countries and institutions because of their low number. Moreover, their proportion with reference to the total number of publications varies across countries (Rodríguez-Navarro 2012), which prevents its calculation. Thus, a reliable indicator should be calculated from "normal publications" but should correlate with the number of landmark publications (Rodríguez-Navarro and Brito 2019). Consequently, the most important step in the proposal of a bibliometric indicator based on a large number of papers is its validation (Harnad 2008), but, unfortunately, this validation requires a standard of comparison that is unclear. In other words, the generation of indicators for research assessment is easier than their validation.

It has already been mentioned above that research assessments with experts is complex and onerous (Martin 2011; Régibeau and Rockett 2016), but in the absence of well-founded bibliometric indicators this is the only reliable method. Therefore, the research institutions in several countries are evaluated by experts (Wouters et al. 2015). When these peer review assessments are well performed, they not only give a solution to a public policy requirement, but such assessments provide priceless information for the validation of bibliometric indicators (Harnad 2009). Certainly, peer review is not guaranteed to be fault-free, but "the natural way to test the validity of metrics is against peer review" (Harnad 2008, p.105); conversely, these validated indicators eventually could substitute for the peer review. However, although most of the peer judgments that are used to validate bibliometric indicators come from the assessments of institutions and research groups, validations can also be based on other types of expert decisions (e.g. Bornmann and Marx 2015; Dunaiski et al. 2016).

Among a notable number of research assessments of institutions based on peer review (Wouters et al. 2015) those carried out in the UK for almost 30 years—the Research Assessment Exercise (RAE) and Research Excellence Framework (REF) —are the most firmly established and most extensively studied. The last research assessment of UK universities, REF, has given rise to an extensive and well-documented study, *The Metric Tide*, about the possible use of bibliometric indicators to substitute for peer review (Wilsdon et al. 2015), and this study has been recently further complemented (Traag and Waltman 2019). Most of these studies address the important question of whether the bibliometric and REF evaluations of the papers presented to REF are correlated. However, these studies do not address the subsequent question of whether a bibliometric indicator that is not based on the REF submitted papers might substitute for the REF assessments. This indicator could perform the evaluation without requiring university applications.

To answer this question, we must return to the aforementioned very low proportion of breakthrough publications and their almost impossible counting in most countries and institutions. A mathematical alternative to this problem of counting is to calculate their probability or expected frequency, which is possible from the power law that holds in the distribution of papers in global percentiles (Rodríguez-Navarro and Brito 2019). From this power law, the $e_p$ index, which is an evaluative designed transformation of the exponent, can be used to calculate probabilities and frequencies of papers at any global percentile.

In the described scenario, the present study was designed with two overlapping aims: to validate percentile indicators calculated from the $e_p$ index and to investigate whether any of these indicators could be used to eventually substitute for the peer-review-based REF evaluations in a like-for-like manner.

## 2. Validation of percentile indicators against the UK REF results

Among the many indicators studied by Wilsdon et al. (2015), Traag and Waltman (2019) use a percentile indicator. Percentile distributions have been widely used for many years in almost all social and technological fields, from medicine (e.g. Acheson 1973) to economics (e.g. Gallman 1969), and metric data in these fields are similar to citations in scientific publications. For example, in analogy with income distributions, "instead of individuals we have scientific articles, and instead of dollars we have citations" (Albarrán et al. 2011, p. 325). An obvious advantage for the use of percentile distributions of citations is that they produce results that are normalized, eliminating the great differences in citations that occur across scientific fields, which otherwise would make it impossible to make field comparisons (Waltman and van-Eck 2013). Furthermore, percentile-based normalization does not have the flaws of normalizing approaches based on arithmetic averages (Bornmann et al. 2013b), and their calculation, opportunity, and limits of use are well established (Bornmann et al. 2013a; Waltman and Schreiber 2013).

The REF assesses three types of criteria: outputs, impact, and environment (REF2014 2011), and peers must rate publications in four starred levels (4*, 3*, 2*, and 1*), which are described as world-leading, internationally excellent, internationally recognised, and nationally recognised, respectively. Of the three evaluated criteria by the REF, the outputs criterion has the highest weight and the results of this criterion are the ones that can be compared to the results obtained with bibliometric indicators.

The study by Wilsdon et al. (2015) compares the scores of peer-reviewed outputs with bibliometric indicators, reporting that the "correlation analysis of the REF2014 results at output-by-author level (Supplementary Report II) has shown that individual metrics give significantly different outcomes from the REF peer review process, and therefore cannot provide a like-for-like replacement for REF peer review" (p. ix).

In contrast, the study by Traag and Waltman (2019) finds a high level of agreement between the top 10% indicator and the scores of peer reviews. It also finds that comparisons are improved if four conditions are fulfilled: (i) comparisons are made at institutional level instead of at output-by-author level; (ii) size-independent indicators are used instead of size-dependent indicators; (iii) correlations are complemented with other types of comparisons; and (iv) taking into account that peer review has a certain level of uncertainty. In their study the percentage of submitted publications that belong to the top 10% of most cited publications ($PP_{top\ 10\%}$ in the Leiden Ranking notation) is compared to the PP(4*), which is the percentage of 4-star-rated papers.

## 3. Previous considerations and aim of this study

### 3.1. General considerations about the REF peer review

The study by Traag and Waltman (2019) describes methods and provides results that strongly support that with certain restrictions the proportion of outputs rated 4-star or world leader class by peer review is in agreement with the $PP_{top\ 10\%}$ indicator. However, to go a step further, towards the use of a bibliometric indicator that makes applications unnecessary, the indicator must be based on the total number of publications from the university instead of on a sample of them. Therefore, in the comparison with the REF results four considerations have to be taken into account.

(i) An important constrain of the REF peer review is that it has to be performed on samples and not on the total number of published papers to limit costs and administrative burden. In the REF the limit was "four outputs listed against each member of staff entered in the exercise" (Wilsdon et al. 2015, p. 119). These REF samples are not random samples but samples containing the outputs that the staff members consider their top outputs. In a six-year evaluation, 22% of the outputs submitted were rated 4-star (Wilsdon et al. 2015, p. 122). Taking this figure, it can be guessed that in medium-level universities it is unlikely that even top researchers have more than two or three 4-star-level publications, which implies that the outputs sample submitted for evaluation may include all 4-star-level publications of the university. In contrast, in the most active universities, in big research groups some staff members may have more than four 4-star level publications, which implies that the sample will contain only a fraction of all 4-star level publications of the university. In consequence, these universities will be sub-evaluated compared to medium-level universities.

(ii) The top percentile of the citation distribution to be used as bibliometric indicator has to be defined. It is intuitive that peer review assessments based on top research publications, i.e. world leader (4-star) class, is conceptually equivalent to top percentiles in the distribution of world publications, but the corresponding percentile is absolutely unknown. Traag and Waltman (2019) use the $PP_{top\ 10\%}$, which seems reasonable but not necessarily accurate. For example, attending to the study by Tijssen et al. (2002), the $PP_{top\ 1\%}$ would have also been reasonable.

(iii) Traag and Waltman (2019) make the important caveat that "correlations between metrics and peer review may not be the most informative measure of agreement" (p. 2). Therefore, they used a test based on median differences. This is an important step forward, but to demonstrate that a like-for-like substitution can be achieved it must be demonstrated that the ratio between the peer review numerical assessment and the bibliometric indicator is 1.0. Certainly, neither peer review (Harnad 2008; Traag and Waltman 2019) nor bibliometric indicators can be expected to provide a perfect measure, which implies that individual ratios will not all be 1.0, but the mean should be close to 1.0 and the standard deviation should be low if a like-for-like substitution is pursued. This consideration can be used in the search for the appropriate top percentile indicator (ii).

(iv) The last consideration regarding peer review is that it does not provide a level that can be compared to universities in other countries; the results of the evaluation are only for internal use. This occurs because a peer-established "world leader level" is a subjective concept that has no external reference. For example, if 30% of the research outputs of University A and 20% of them in University B are rated 4-star or world leader, these two universities can be compared between themselves but neither of them can be compared to USA universities.

## 3.2. Collaborative studies hinder both peer and bibliometric evaluations

Research evaluation using methods based on either bibliometrics or peer review has pros and cons (see a review in Wouters et al. 2015), but a singular problem arises when both peer review and bibliometrics are unable to perform the evaluation reliably. This occurs with publications where the number of participant authors and institutions is so high that the assessment of the actual merit of each institution or author is practically impossible.

The field of scientific collaboration has been extensively studied from many points of view (reviewed by Sonnenwald 2007), including unethical practices (Cronin 2001), which are deliberately ignored here. The number of publications with more than one institution has increased over the last 50 years (Wuchty et al. 2007); from a bibliometric point of view it is known that collaborations have the effect of increasing the number of citations (Persson et al. 2004), due—but not exclusively—to self-citation (Wuchty et al. 2007). This increase in citations might be difficult to interpret for the evaluation of the paper, but the actual problem arises when individual merits have to be assigned to either authors or institutions. When the number of institutions is low—e.g. up to three or four—assigning the real merit to each one of the participant institutions might be difficult but is not an impossible task for experts in the field, and fractional counting (Waltman and van-Eck 2015) may be a reasonably bibliometric solution. In this case, even full counting might not be a distorting solution.

However, the important issue in the evaluations of collaborative publications is that in certain research fields the number of participant institutions can be hundreds (Birnholtz 2006; King 2012). These consortia are typical in the fields of particle physics, genome sequencing, and clinical trials, and a reliable evaluation of the merit of each participant institution or researcher in these papers may be practically impossible. If the proportion of these publications in both the global and institutional production is small, their biasing effect will be small and irrelevant. In contrast, if the proportion is high, reliable individual assessments may be impossible. The use of formal methods of weighting (Rossi et al. 2019) is a statistical solution, but that does not distinguish individual merits.

### 3.3. Aim of the present study

This study was designed to find bibliometric indicators that could substitute for peer assessments in a like-for-like manner, which implies that if the indicator is found, it is simultaneously validated. For this purpose, this study is based on the outputs results of REF. The notion of the existence of a validatable bibliometric indicator that is calculated from the total number of publications recorded in databases seems plausible in research fields in which most of their research results are communicated through journal articles. In REF, in natural and formal sciences, and in technologies practically all submitted outputs are journal articles. In some social sciences, such as Economics and Econometrics, and Business and Management Studies not all, but a large proportion of outputs (> 90%) are journal articles (Wilsdon et al. 2015, p. 154).

Traag and Waltman (2019) have demonstrated that when considering exclusively the outputs submitted for peer assessment, the level of agreement between the $PP_{top\ 10\%}$ and $PP(4^*)$ of universities is very high. Pearson correlation coefficient varied depending on the field of research but in most cases was higher or slightly lower than 0.8; these results clearly establish that a percentile indicator can be the ideal bibliometric indicator that allows a like-for-like substitution for peer review. To go a step beyond this idea, our study pursued three specific aims:

1. To determine the percentile indicator ($PP_{top\ x\%}$) that corresponds to the 4-star level of peer review, fulfilling the condition that the $PP_{top\ x\%}/PP(4^*)$ ratio is 1.0, which implies that it is a like-for-like substitute.

2. To calculate individual deviations of the $PP_{top\ x\%}/PP(4^*)$ ratio from 1.0. Firstly, to estimate whether the $PP_{top\ x\%}$ indicator may be a like-for-like substitute for peer review, and, secondly, if this substitution is possible, to identify cases of high deviations that can be investigated by experts in the field.

3. To discuss the advantages and limits of using a PP$_{top\ x\%}$ indicator for the research assessments of institutions as a substitute for peer reviews.

## 4. Methods and data

To investigate the most convenient PP$_{top\ x\%}$ indicator for the purposes just stated, we used the percentile-based double rank analysis of citation frequencies to calculate the $e_p$ index (Rodríguez-Navarro and Brito 2018). When the research performance of an institution of country is coincident with the global average, the $e_p$ index values 0.1; maximum values of the $e_p$ index are around 0.20−0.25. To calculate the $e_p$ index, we counted the number of publications in the global percentiles 7, 10, 14, 20, 27, and 35 of the research fields, and fitted the data to a power law (Rodríguez-Navarro and Brito 2019); as in REF, we used full counting for each publication authored by several universities. The publication window was one year. In universities, the interannual variability of the $e_p$ index is notable in some cases. To overcome this annual variability we used the mean of four years 2009–12. For this purpose, for each year, we calculated the PP$_{top}$ values for each one of the aforementioned percentiles—percentage of the papers from the university in each global percentile. Next, we calculated the means of the four PP$_{top}$ values for these percentiles and these means were then used to fit the power law and calculate the $e_p$ index of the university. The one-year publication window raises a problem in the analysis of some universities because in many fields of research, the total number of publications is low and our limit for an accurate fitting is about 80–120 publications. With this number of publications, the goodness of fit was variable—better fits seem to be associated to higher $e_p$ index values. In the universities presented in this study the fits to the power law showed $R^2$ and $p$ values calculated by using the $X^2$ statistics (Press et al 1989) that were higher than 0.99.

The PP$_{top\ x\%}$ indicators were calculated with the following formula (Rodríguez-Navarro and Brito 2019):

$$PP_{top\ x\%} = 100 \cdot e_p^{(2-\lg x)} \qquad (1)$$

by giving values to *x* it is possible to select the *x* value that makes the $PP_{top\ x\%}/PP(4^*)$ ratio equal 1.0. Using the same formula it can be calculated the $PP_{top\ 0.01\%}$. This indicator is 100-times the probability that one paper of the university is in the 0.01 percentile, which is a reasonable indicator of research landmark (Bornmann et al. 2018) even at the level of a Nobel Prize (Brito and Rodríguez-Navarro 2018a).

The REF reports the results of 36 units of assessment (UOA; REF2014 2011) and we have studied four of these units: Chemistry (# 8); Physics (# 9); and Economics & Econometrics (# 18) joined to Business & Management Studies (# 19). To obtain the bibliometric data, we used the Web of Science (WoS), making the searchers using the Advanced Search feature and the research areas (SU=) of Chemistry, Physics, and Business & Economics joined to Operations Research & Management Science, which were matched with the UOAs above, respectively. For the Chemistry and Physics research areas, we used the database Science Citation Index Expanded; for the Business & Economics and Operations Research & Management Science, we used two databases, the Science Citation Index Expanded and the Social Sciences Citation Index. To retrieve the papers published by each university, we used the Organization-Enhanced (OG=) tool of the database. We restricted the search to only "articles" because review papers receive more citations than original papers, but it is unlikely that they receive a better qualification than original papers in peer review assessments. Furthermore, review papers may distort citation distribution (Brito and Rodríguez-Navarro 2019). We studied the articles published in each year from 2009 to 2012, recording the number of citations up to the day of the search. Because percentile analyses require the analyses of world and institution publications, we obtained the world and institution numbers of citations on the same day.

The REF outputs results referred to universities and OUAs were retrieved at https://results.ref.ac.uk, restricting our study to the 4-star level. Our method can be equally applied to the addition of the 4- and 3-star levels. However, in most universities we have studied, the percentage of joined outputs in both levels exceeds the 90% (considering all universities the percentage is 72%), which indicates that the outputs presented for evaluation will be considerably less than the total production of putative 4- and 3-star level papers of the university. This observation precludes comparisons between REF outputs data and the $PP_{top\ x\%}$ that correspond to the joint of 4- and 3-star-rated papers because the REF sample is incomplete.

As already explained, the $PP_{top\ x\%}$ indicators were calculated from the papers retrieved from the WoS database and correspond to the whole production of the university. In contrast, the REF 4-star outputs data is a percentage of the research outputs submitted. Therefore, to calculate the $PP_{top\ x\%}/ PP(4^*)$ ratio it was necessary to express the $PP(4^*)$ results as percentages of the whole production. For this purpose, we assumed that the REF recorded 4-star outputs make up the total number of publications of this level of the university. Under this assumption, the $PP(4^*)$ indicator, which is the percentage of 4-star-rated publications in the whole production, was calculated from the number of submitted outputs, the percentage of 4-star outputs, and the number of papers retrieved from WoS. We first calculated the number of 4-star-rated outputs, this number was referred to the total number of papers retrieved from the WoS database, and the ratio was expressed as a percentage. Thus, although for consistency we keep the $PP(4^*)$ notation of Traag and Waltman (2019), their and our parameters are not identical: theirs is a percentage referred to the number of submitted outputs and ours a percentage referred to the total number of publications.

For comparisons with external universities, we analysed the publications of the Massachusetts Institute of Technology (MIT) in the WoS research areas of Chemistry and Physics, and of the Princeton University in the research areas of Business & Economics joined to Operations Research & Management Science. To calculate the indicators for these universities, we proceeded as for the UK universities.

## 5. Results

### 5.1. Chemistry

The REF lists 35 universities in the UOA of Chemistry (# 8); Table 1 records these universities, including the number of outputs submitted and percentage of these outputs

Table 1. Summary of the REF2014 university outputs in the unit of assessment of Chemistry that meet the 4-star standard and calculated PP(4*) indicator

| University | Outputs submitted | REF 4-star (%) | WoS papers 2008-2013 | PP(4*) (%)[a] |
|---|---|---|---|---|
| University of Bath | 122 | 18.9 | 1029 | 2.24 |
| University of Birmingham | 109 | 13.8 | 934 | 1.61 |
| University of Bristol | 236 | 28.0 | 1336 | 4.95 |
| University of Cambridge | 229 | 46.5 | 3045 | 3.53 |
| University of Durham | 152 | 23.0 | 1071 | 3.26 |
| University of East Anglia | 65 | 32.3 | 724 | 2.90 |
| University of Greenwich | 59 | 3.4 | 148 | 1.36 |
| University of Huddersfield | 62 | 4.8 | 166 | 1.79 |
| University of Hull | 94 | 9.6 | 357 | 2.53 |
| Imperial College London | 217 | 20.7 | 2398 | 1.87 |
| University of Kent | 57 | 14.0 | 190 | 4.20 |
| Lancaster University | 32 | 25.0 | 181 | 4.42 |
| University of Leeds | 125 | 13.6 | 1169 | 1.45 |
| University of Leicester | 78 | 6.4 | 273 | 1.83 |
| University of Liverpool | 119 | 44.5 | 865 | 6.12 |
| University College London | 248 | 22.2 | 2034 | 2.71 |
| Loughborough University | 90 | 1.1 | 522 | 0.19 |
| University of Manchester | 207 | 20.8 | 2239 | 1.92 |
| Newcastle University | 95 | 4.2 | 672 | 0.59 |
| University of Nottingham | 154 | 17.5 | 1463 | 1.84 |
| University of Oxford | 314 | 38.2 | 3315 | 3.62 |
| Queen Mary University of London | 45 | 31.1 | 373 | 3.75 |
| University of Reading | 88 | 12.5 | 697 | 1.58 |
| University of Sheffield | 112 | 23.2 | 1119 | 2.32 |
| University of Southampton | 159 | 26.4 | 1218 | 3.45 |
| University of Sussex | 65 | 16.2 | 256 | 4.11 |
| University of Warwick | 134 | 29.1 | 1093 | 3.57 |
| University of York | 191 | 24.1 | 742 | 6.20 |
| University of Aberdeen | 78 | 9.0 | 343 | 2.05 |
| Universities of Edinburg and St Andrews | 146 | 22.7 | 2482 | 1.34 |
| Universities of Glasgow and Strathclyde | 120 | 17.4 | 1741 | 1.20 |
| Heriot-Watt University | 113 | 15.0 | 529 | 3.20 |
| Bangor University | 40 | 5.0 | 136 | 1.47 |
| Cardiff University | 103 | 22.3 | 1017 | 2.26 |
| Queen's University Belfast | 138 | 5.1 | 850 | 0.83 |

[a] Percent of publications that meet the 4-star standard with reference to the total number of publications retrieved for the WoS research area of Chemistry

that were rated 4-star. The last two columns of Table 1 show the number of papers retrieved from the WoS for the 2008–13 period in the WoS research area of Chemistry, and the percentage that the number of the 4-star level outputs represents in the total number of WoS publications.

Table 2. Substitution of a percentile indicator for peer review in the research field of chemistry. Comparison of the bibliometric indicator $PP_{top\ 2.8\%}$ with the proportion of 4-star rated outputs by peer-review in the UOA of Chemistry in REF2014, and values of the $PP_{top\ 0.01\%}$ indicator

| University | $e_p$ index | PP(4*) | $PP_{top\ 2.8\%}$ | $PP_{top\ 0.01\%}$ | Ratio $PP_{top\ 2.8\%}$/ PP(4*) |
|---|---|---|---|---|---|
| Imperial College London | 0.166 | 1.87 | 6.17 | 0.0766 | **3.30** |
| Newcastle University | 0.075 | 0.59 | 1.79 | 0.0032 | **3.02** |
| Universities St Andrews and Edinburg | 0.118 | 1.34 | 3.60 | 0.0192 | **2.70** |
| University of Cambridge | 0.185 | 3.53 | 7.30 | 0.1180 | **2.07** |
| University of Bath | 0.134 | 2.24 | 4.40 | 0.0321 | **1.96** |
| University of Manchester | 0.109 | 1.92 | 3.22 | 0.0143 | 1.67 |
| University of Hull | 0.120 | 2.53 | 3.72 | 0.0207 | 1.47 |
| University of Nottingham | 0.096 | 1.84 | 2.61 | 0.0083 | 1.42 |
| University of Aberdeen | 0.097 | 2.05 | 2.69 | 0.0090 | 1.31 |
| University of Birmingham | 0.074 | 1.61 | 1.77 | 0.0031 | 1.10 |
| University of Leeds | 0.070 | 1.45 | 1.59 | 0.0023 | 1.10 |
| Cardiff University | 0.091 | 2.26 | 2.41 | 0.0068 | 1.07 |
| University of York | 0.175 | 6.20 | 6.66 | 0.0933 | 1.07 |
| Universities Strathclyde and Glasgow | 0.059 | 1.20 | 1.22 | 0.0012 | 1.02 |
| University College London | 0.096 | 2.71 | 2.65 | 0.0086 | 0.98 |
| University of Oxford | 0.115 | 3.62 | 3.49 | 0.0177 | 0.97 |
| Durham University | 0.108 | 3.26 | 3.14 | 0.0134 | 0.96 |
| University of Sheffield | 0.082 | 2.32 | 2.06 | 0.0045 | 0.89 |
| University of Southampton | 0.106 | 3.45 | 3.05 | 0.0125 | 0.88 |
| University of Reading | 0.062 | 1.58 | 1.34 | 0.0015 | 0.85 |
| Queen Mary University of London | 0.108 | 3.75 | 3.16 | 0.0136 | 0.84 |
| University of Warwick | 0.103 | 3.57 | 2.92 | 0.0112 | 0.82 |
| University of East Anglia | 0.089 | 2.90 | 2.35 | 0.0064 | 0.81 |
| University of Liverpool | 0.145 | 6.12 | 5.01 | 0.0447 | 0.82 |
| University of Leicester | 0.064 | 1.83 | 1.41 | 0.0017 | 0.77 |
| University of Bristol | 0.111 | 4.95 | 3.27 | 0.0149 | 0.66 |
| | | | | | |
| Mean ratio excluding the top five ratios | | | | | 1.02 |
| SD | | | | | 0.26 |
| | | | | | |
| Massachusetts Institute of Technology | 0.247 | | 11.89 | 0.3751 | |

Next, we calculated the $e_p$ index for these universities, as explained in Section 3. Excluding the universities in which the number of publications was too low, the number of universities was reduced to 26. The next step was to give values to *x* in Eq. (1) in order that the mean of the $PP_{top\ x\%}/PP(4^*)$ ratios of the 26 universities was as close as possible to 1.0. The 1.9 percentile fulfils this condition (mean = 1.004). However, at this percentile, and at any other, five universities deviate from the trend of the other 21: Imperial College of London; Newcastle University; the joint submission of the Universities of Edinburgh and St Andrews; the University of Cambridge; and the University of Bath. Therefore, they were omitted from the calculation of the 4-star-rated equivalent percentile in order to study their deviations independently. Excluding these universities, the 2.8 percentile fulfils the condition (the mean ratio was 1.02 and SD = 0.26; a mean ratio closer to 1.0 can be obtained using a percentile with two decimal figures). Table 2 records the calculated $e_p$ index values and $PP_{top\ 2.8\%}/PP(4^*)$ ratios for the 26 universities under study.

Table 2 also shows the $PP_{top\ 0.01\%}$ values for each university. As mentioned before, this indicator is 100-times the probability that a random paper of the university is in the 0.01 percentile, which is a reasonable indicator of research excellence (Bornmann et al. 2018; Brito and Rodríguez-Navarro 2018a). Considering this indicator and the $e_p$ index, the research competitiveness of the MIT is much higher than in the UK universities.

**5.2. Economics and business**

Because the UOAs in this area and the WoS research areas were not coincident, we joined the REF results in Economics and Econometrics (REF, UOA # 18) and in Business and Management Studies (REF, UOA #19), and the WoS research areas of Business & Economics and Operations Research & Management Science. We compared the two joint UOA areas with the two joint WoS areas.

REF lists 98 universities with outputs in the UOAs #18 and #19; Table 3 records these universities including the number of outputs submitted to each UOA and percentage of the outputs that were rated 4-star. The last two columns of Table 3 show the number of papers retrieved from the WoS for the 2008–13 period in the research areas of Business & Economics and Operations Research & Management Science, and the percentage that the number of the 4-star rated outputs in the UOAs #18 and #19 represents in the total number of WoS publications.

Table 3. Summary of the REF2014 university outputs in the unit of assessment of Economics & Econometrics joined to Business & Management Studies that meet the 4-star standard and calculated PP(4*) indicator

| University | WoS 2008-2013 | OUA | | | | # 18 and 19 |
| | | #18 | | #19 | | |
| | | submitted | 4-star (%) | submitted | 4-star (%) | PP(4*) (%)[a] |
|---|---|---|---|---|---|---|
| Anglia Ruskin University | 16 | 97 | 10.3 | 44 | 6.8 | 81.14 |
| Aston University | 349 | | | 174 | 21.3 | 10.62 |
| University of Bath | 483 | | | 207 | 27.5 | 11.79 |
| University of Bedfordshire | 42 | | | 47 | 10.6 | 11.86 |
| Birkbeck College | 131 | | | 103 | 12.6 | 9.91 |
| University of Birmingham | 554 | | | 204 | 17.2 | 6.33 |
| Birmingham City University | 15 | 79 | 7.6 | 17 | 5.9 | 46.71 |
| Bournemouth University | 119 | | | 65 | 4.6 | 2.51 |
| University of Bradford | 178 | | | 71 | 18.3 | 7.30 |
| University of Brighton | 49 | | | 69 | 17.4 | 24.50 |
| University of Bristol | 275 | 63 | 22.4 | 85 | 15.3 | 9.86 |
| Brunel University London | 435 | 102 | 2 | 228 | 11.4 | 6.44 |
| University of Cambridge | 1143 | 99 | 54.5 | 163 | 43.6 | 10.94 |
| University of Central Lancashire | 71 | | | 63 | 3.2 | 2.84 |
| University of Chester | 13 | | | 23 | 4.3 | 7.61 |
| City University London | 605 | 54 | 16.7 | 330 | 36.6 | 21.45 |
| Coventry University | 73 | | | 66 | 4.5 | 4.07 |
| Cranfield University | 367 | | | 154 | 14.3 | 6.00 |
| De Montfort University | 110 | | | 82 | 8.5 | 6.34 |
| University of Derby | 9 | | | 35 | 0 | 0.00 |
| University of Durham | 315 | | | 179 | 25.7 | 14.60 |
| University of East Anglia | 451 | 49 | 20.4 | 75 | 30.7 | 7.32 |
| University of East London | 46 | | | 16 | 0 | 0.00 |
| University of Essex | 440 | 113 | 29.2 | 165 | 17 | 13.87 |
| University of Exeter | 396 | 83 | 13.3 | 171 | 17.5 | 10.34 |
| University of Greenwich | 88 | | | 113 | 7.1 | 9.12 |
| University of Hertfordshire | 88 | | | 67 | 7.5 | 5.71 |
| University of Huddersfield | 39 | | | 67 | 6 | 10.31 |
| University of Hull | 204 | | | 157 | 9.6 | 7.39 |
| Imperial College London | 692 | | | 204 | 48.5 | 14.30 |
| Keele University | 66 | | | 63 | 9.5 | 9.07 |
| University of Kent | 340 | 79 | 2.5 | 158 | 17.7 | 8.81 |
| King's College London | 292 | | | 147 | 24.5 | 12.33 |

| Institution | | | | | | |
|---|---:|---:|---:|---:|---:|---:|
| Kingston University | 131 | | | 107 | 16.8 | 13.72 |
| Lancaster University | 634 | | | 461 | 24.9 | 18.11 |
| University of Leeds | 583 | | | 262 | 22.1 | 9.93 |
| Leeds Beckett University | 50 | | | 72 | 1.4 | 2.02 |
| University of Leicester | 378 | 80 | 18.8 | 218 | 14.5 | 12.34 |
| University of Lincoln | 44 | | | 28 | 10.3 | 6.55 |
| University of Liverpool | 268 | | | 156 | 8.9 | 5.18 |
| University College London | 721 | 142 | 69.7 | 40 | 55 | 16.78 |
| London Business School | 486 | | | 356 | 55.3 | 40.51 |
| London School of Economics and Political Science | 1500 | 183 | 56.3 | 296 | 47.6 | 16.26 |
| London Metropolitan University | 135 | | | 13 | 7.7 | 0.74 |
| London South Bank University | 19 | | | 35 | 2.9 | 5.34 |
| Loughborough University | 521 | | | 230 | 22.2 | 9.80 |
| University of Manchester | 1267 | 114 | 11.4 | 456 | 20.8 | 8.51 |
| Manchester Metropolitan University | 122 | | | 80 | 5 | 3.28 |
| Middlesex University | 181 | | | 170 | 11.6 | 10.90 |
| Newcastle University | 342 | | | 231 | 18.6 | 12.56 |
| University of Northampton | 15 | | | 32 | 0 | 0.00 |
| University of Northumbria at Newcastle | 84 | | | 76 | 5.3 | 4.80 |
| University of Nottingham | 1183 | 127 | 19.7 | 321 | 16.2 | 6.51 |
| Nottingham Trent University | 136 | | | 98 | 13.3 | 9.58 |
| Open University | 165 | | | 74 | 13.5 | 6.05 |
| School of Oriental and African Studies | 121 | | | 91 | 8.8 | 6.62 |
| University of Oxford | 1404 | 242 | 42.6 | 156 | 44.2 | 12.25 |
| Oxford Brookes University | 121 | | | 85 | 9.4 | 6.60 |
| University of Plymouth | 124 | | | 125 | 9.6 | 9.68 |
| University of Portsmouth | 135 | | | 156 | 7.7 | 8.90 |
| Queen Mary University of London | 225 | 94 | 20.2 | 111 | 19.8 | 18.21 |
| University of Reading | 397 | | | 139 | 18.7 | 6.55 |
| Roehampton University | 16 | | | 19 | 0 | 0.00 |
| Royal Holloway, University of London | 223 | 51 | 35.5 | 168 | 23.8 | 26.05 |
| University of Salford | 177 | | | 72 | 5.6 | 2.28 |
| University of Sheffield | 548 | 50 | 8 | 119 | 23.5 | 5.83 |
| Sheffield Hallam University | 52 | | | 28 | 3.6 | 1.94 |
| University of Southampton | 521 | 82 | 22 | 124 | 12.1 | 6.34 |
| Staffordshire University | 38 | | | 33 | 3 | 2.61 |
| University of Sunderland | 9 | | | 16 | 0 | 0.00 |
| University of Surrey | 301 | 71 | 26.8 | 148 | 16.9 | 14.63 |
| University of Sussex | 362 | 54 | 14.8 | 139 | 18 | 9.12 |
| Teesside University | 27 | | | 21 | 9.5 | 7.39 |
| University of Warwick | 1017 | 136 | 42.6 | 374 | 38.2 | 19.74 |
| University of the West of England, Bristol | 192 | | | 131 | 9.9 | 6.75 |
| University of Westminster | 111 | | | 76 | 7.9 | 5.41 |
| University of Wolverhampton | 29 | | | 37 | 2.7 | 3.44 |
| University of Worcester | 5 | | | 28 | 0 | 0.00 |
| University of York | 498 | 104 | 14.4 | 81 | 17.3 | 5.82 |
| York St John University | 2 | | | 24 | 0 | 0.00 |

| University | | | | | | |
|---|---|---|---|---|---|---|
| University of Aberdeen | 285 | 63 | 4.8 | 53 | 17 | 4.22 |
| University of Dundee | 106 | | | 72 | 5.6 | 3.80 |
| University of Edinburgh | 422 | 55 | 30.9 | 166 | 21.1 | 12.33 |
| Edinburgh Napier University | 59 | | | 44 | 11.1 | 8.28 |
| University of Glasgow | 415 | 83 | 18.1 | 131 | 18.3 | 9.40 |
| Glasgow Caledonian University | 69 | | | 65 | 4.6 | 4.33 |
| Heriot-Watt University | 203 | | | 119 | 8.4 | 4.92 |
| Robert Gordon University | 46 | | | 31 | 9.7 | 6.54 |
| University of St Andrews | 231 | 51 | 23.5 | 74 | 24.3 | 12.97 |
| University of Stirling | 239 | | | 137 | 15.3 | 8.77 |
| University of Strathclyde | 532 | | | 309 | 18.8 | 10.92 |
| University of the West of Scotland | 23 | | | 35 | 5.7 | 8.67 |
| Aberystwyth University | 69 | | | 52 | 7.7 | 5.80 |
| Bangor University | 163 | | | 105 | 19 | 12.24 |
| Cardiff University | 709 | | | 272 | 27.2 | 10.43 |
| Swansea University | 184 | | | 101 | 8.9 | 4.89 |
| Queen's University Belfast | 280 | | | 184 | 20.7 | 13.60 |
| University of Ulster | 157 | | | 95 | 23.2 | 14.04 |

[a] Percent of publications that meet the 4-star standard with reference to the total number of publications retrieved for the WoS research areas of Business & Economics and Operations Research & Management Science

Next, we calculated the $e_p$ index for these universities, excluding those in which the number of publications was too low (Section 3). These exclusions reduced the number of universities to 15. This significant reduction occurs because the number of WoS papers in most universities in the UOAs #18 and #19 is low (Table 3). In contrast, the goodness of fits of data to the power law was high even with numbers of papers below 100.

As above, the next step was to give values to $x$ in Eq. (1) in order that that the mean of the $PP_{top\ x\%}/PP(4*)$ ratios of the 15 universities was as close as possible to 1.0. The 9.0 percentile fulfils this condition (mean = 1.03), but the data showed a notable variability (SD = 0.43), which, in contrast to Chemistry, did not occur because a few universities deviated from the trend of the others. Notably, in the two universities with the highest number of Nobel Laureates in the field of Economic Sciences, Cambridge and the London School of Economics and Political Sciences, the $PP_{top\ 9.0\%}/PP(4*)$ ratios were 1.21 and 0.70, which did not deviate very much from the mean value of 1.0. Between these two values of the ratio there are six universities (Table 4); outside these values there are eight universities, four with higher and four with lower ratios. This symmetry around the central value of 1.0 demonstrates that there are no individual deviations from a general trend.

Table 4. Substitution of a percentile indicator for peer review in the research field of economics and business. Comparison of the bibliometric indicator PP$_{top\ 9.0\%}$ with the proportion of 4-star rated outputs by peer-review in the UOAs of Economics & Econometrics joined to Business & Management Studies in REF2014, and values of the PP$_{top\ 0.01\%}$ indicator

| University | $e_p$ index | PP(4*) | PP$_{top\ 9\%}$ | PP$_{top\ 0.01\%}$ | Ratio PP$_{top\ \%\ 9\%}$/PP(4*) |
|---|---|---|---|---|---|
| University of Nottingham | 0.132 | 6.51 | 12.02 | 0.033 | 1.85 |
| University of Leeds | 0.197 | 9.93 | 18.33 | 0.167 | 1.85 |
| University of York | 0.100 | 5.82 | 9.01 | 0.011 | 1.55 |
| University of Sheffield | 0.083 | 5.83 | 7.41 | 0.005 | 1.27 |
| University of Cambridge | 0.144 | 10.94 | 13.21 | 0.048 | 1.21 |
| University of Birmingham | 0.078 | 6.33 | 6.98 | 0.004 | 1.10 |
| Imperial College London | 0.163 | 14.30 | 14.98 | 0.077 | 1.05 |
| University of Bath | 0.134 | 11.79 | 12.20 | 0.035 | 1.04 |
| University of Oxford | 0.128 | 12.25 | 11.64 | 0.029 | 0.95 |
| Cardiff University | 0.106 | 10.43 | 9.56 | 0.014 | 0.92 |
| University of Manchester | 0.080 | 8.51 | 7.14 | 0.005 | 0.84 |
| London School of Economics and Political Science | 0.125 | 16.26 | 11.35 | 0.027 | 0.70 |
| University of Strathclyde | 0.085 | 10.92 | 7.55 | 0.006 | 0.69 |
| City University London | 0.130 | 21.45 | 11.87 | 0.032 | 0.55 |
| University of Warwick | 0.117 | 19.74 | 10.63 | 0.021 | 0.54 |
| University College London | 0.076 | 16.78 | 6.78 | 0.004 | 0.40 |
|  |  |  |  |  |  |
| Mean ratio |  |  |  |  | 1.03 |
| SD |  |  |  |  | 0.43 |
|  |  |  |  |  |  |
| Princeton University | 0.238 |  | 22.31 | 0.322 |  |

Table 4 also shows the PP$_{top\ 0.01\%}$ values for each university. Considering this indicator and the $e_p$ index, the research competitiveness of Princeton University is much higher than in the UK universities.

### 5.3. Physics

The REF lists 40 universities with outputs in the UOA of Physics (# 9); Table 5 records these universities including the number of outputs submitted, the percentage of 4-star-rated outputs, and the number of publications retrieved from the WoS.

As in the previous UOAs, the next step was to calculate the $e_p$ index with the data obtained from the WoS. For this purpose, some universities could not be studied because the low number of publications in some or in all years of the study was too low.

Table 5. Summary of the REF2014 university outputs in the unit of assessment of Physics that meet the 4-star standard and calculated PP(4*) indicator

| University | Outputs submitted | REF 4-star (%) | WoS 2008-2013 | PP(4*) (%)[a] |
|---|---|---|---|---|
| University of Bath | 84 | 15.5 | 537 | 2.42 |
| University of Birmingham | 157 | 22.9 | 1604 | 2.24 |
| University of Bristol | 191 | 18.8 | 2208 | 1.63 |
| University of Cambridge | 535 | 23.9 | 6043 | 2.12 |
| University of Central Lancashire | 84 | 9.5 | 44 | 18.14 |
| University of Durham | 293 | 21.8 | 1412 | 4.52 |
| University of Exeter | 146 | 21.9 | 505 | 6.33 |
| University of Hertfordshire | 130 | 8.5 | 49 | 22.55 |
| University of Huddersfield | 42 | 9.5 | 55 | 7.25 |
| Imperial College London | 453 | 23.6 | 4548 | 2.35 |
| Keele University | 43 | 23.3 | 83 | 12.07 |
| University of Kent | 17 | 23.5 | 158 | 2.53 |
| King's College London | 97 | 22.7 | 764 | 2.88 |
| Lancaster University | 134 | 27.6 | 1206 | 3.07 |
| University of Leeds | 88 | 13.6 | 1141 | 1.05 |
| University of Leicester | 200 | 9 | 300 | 6.00 |
| University of Liverpool | 138 | 17.4 | 2167 | 1.11 |
| Liverpool John Moores University | 85 | 22.4 | 53 | 35.92 |
| University College London | 446 | 18.6 | 2972 | 2.79 |
| Loughborough University | 75 | 6.7 | 754 | 0.67 |
| University of Manchester | 256 | 17.6 | 2787 | 1.62 |
| University of Nottingham | 193 | 20.7 | 1403 | 2.85 |
| University of Oxford | 464 | 33.2 | 4911 | 3.14 |
| University of Portsmouth | 51 | 21.6 | 276 | 3.99 |
| Queen Mary University of London | 91 | 23.1 | 987 | 2.13 |
| Royal Holloway, University of London | 101 | 17.8 | 689 | 2.61 |
| University of Sheffield | 110 | 23.6 | 1672 | 1.55 |
| University of Southampton | 120 | 25 | 1849 | 1.62 |
| University of Surrey | 101 | 15.8 | 1031 | 1.55 |
| University of Sussex | 95 | 20 | 667 | 2.85 |
| University of Warwick | 215 | 24.1 | 1808 | 2.87 |
| University of York | 137 | 18.2 | 940 | 2.65 |
| University of Edinburgh + University of St Andrews | 224 | 26.8 | 2922 | 2.05 |
| University of Glasgow | 161 | 13.7 | 1933 | 1.14 |
| Heriot-Watt University | 79 | 21.5 | 751 | 2.26 |
| University of Strathclyde | 111 | 27 | 1055 | 2.84 |
| Aberystwyth University | 47 | 2.1 | 87 | 1.13 |
| Cardiff University | 74 | 21.6 | 665 | 2.40 |
| Swansea University | 82 | 13.4 | 497 | 2.21 |
| Queen's University Belfast | 166 | 25.3 | 930 | 4.52 |

[a] Percent of publications that meet the 4-star standard with reference to the total number of publications retrieved for the WoS research area of Physics

In addition to this difficulty, in the UOA of Physics, we found specific anomalies that we had not observed in either this or many other studies (Brito and Rodríguez-Navarro 2018b; Rodríguez-Navarro and Brito 2018). The first surprising observation was that in some universities with a large number of publications, such as for example the University of Oxford, University of Birmingham, Imperial College of London, or University of Southampton, the distribution of percentiles deviated from a power law and the $e_p$ index could not be calculated. Even in some universities, such as the Universities of Cambridge and Manchester, to fit the power law we had to omit one or two data points, which is very unusual in universities with a large number of publications as in these cases. Because of these problems, we could calculate the $e_p$ index in only 12 universities (Table 6). In general terms, the $e_p$ index of these 12 universities was significantly higher than in Chemistry (compare Tables 2 and 6); only in one case was it lower than 0.1 and the value was 0.093, while in Chemistry the $e_p$ index was lower than 0.1 in 40% of the universities. Furthermore, in Chemistry in only one university, the University of Cambridge, was the $e_p$ index higher than 0.15, while in Physics six universities out of 12 had an $e_p$ index higher than 0.15.

Table 6. Substitution of a percentile indicator for peer review in the research field of physics. Comparison of the bibliometric indicator $PP_{top\ 1.1\%}$ with the proportion of 4-star rated outputs by peer review in the UOA of Physics in REF2014, and values of the $PP_{top\ 0.01\%}$ indicator

| University | $e_p$ index | PP(4*) | $PP_{top\ 1.1\%}$ | $PP_{top\ 0.01\%}$ | Ratio $PP_{top\ 1.1\%}$/PP(4*) |
|---|---|---|---|---|---|
| Loughborough University | 0.096 | 0.67 | 1.02 | 0.009 | 1.52 |
| University of Leeds | 0.119 | 1.05 | 1.55 | 0.020 | 1.48 |
| University of Manchester | 0.146 | 1.62 | 2.30 | 0.045 | 1.42 |
| University of Cambridge | 0.164 | 2.12 | 2.90 | 0.072 | 1.37 |
| Universities of Edinburgh and St Andrews | 0.156 | 2.05 | 2.64 | 0.060 | 1.29 |
| University of Strathclyde | 0.174 | 2.84 | 3.27 | 0.093 | 1.15 |
| Cardiff University | 0.159 | 2.4 | 2.73 | 0.064 | 1.14 |
| University of Sheffield | 0.119 | 1.55 | 1.55 | 0.020 | 1.00 |
| University of Durham | 0.167 | 4.52 | 3.02 | 0.079 | 0.67 |
| Lancaster University | 0.138 | 3.07 | 2.06 | 0.036 | 0.67 |
| University of York | 0.108 | 2.65 | 1.29 | 0.014 | 0.49 |
| University of Nottingham | 0.093 | 2.85 | 0.95 | 0.007 | 0.33 |
| | | | | | |
| Mean ratio | | | | | 1.04 |
| Standard deviation | | | | | 0.41 |
| | | | | | |
| Massachusetts Institute of Technology | 0.217 | | 4.40 | 0.223 | |

The next step was to give values to *x* in Eq. (1) in order that the mean of the $PP_{top\ x\%}/PP(4*)$ ratios of the 16 universities was as close as possible to 1.0. The $PP_{top\ 1.1\%}$ fulfilled this condition, but perhaps consistently with the anomalies observed, the variability of the ratios was very high (mean = 1.04; SD = 0.41). The highest ratio amounted to 1.52, and the lowest 0.33.

Aside from other possible difficulties, the high proportion of hyper-authored papers (Section 3.2) was a notable problem for the assessment in Physics. Table 7 shows the distribution of outputs with multiple authors across the universities evaluated in Physics in REF (excluding universities with a very low number of publications). The proportion of these multi-authored papers varies among universities from no multi-authored papers, such as Heriot-Watt University and University of Durham, to the Royal Holloway University of London in which 80% of the papers were multi-authored (Table 7); in half of the universities the proportion was over 20%. Figure 1 shows the distribution of the number of authors in the WoS publications of the universities recorded in Table 7 in the UOA of Physics and, as a comparison, in the UOA of Chemistry (Table 2)—because we used full counting, some publications were counted several times. Up to 30 authors, the distributions for Chemistry and Physics are very similar, although the number of authors per paper was slightly lower in Physics, mode 3–4, than in Chemistry, mode 5–7, but in Physics there is another series of papers where the number of authors varies from 50 to more than 3,000. These papers show two peaks at 300–800 and about 3,000 authors that correspond to international collaborations. The latter were mainly ATLAS and CMS Collaborations using the Large Hadron Collinder at CERN; collaborations with 300–800 authors were diverse, among which LHCb and CDF Collaborations were the most frequent, the former working at CERN and the latter working at Fermilab.

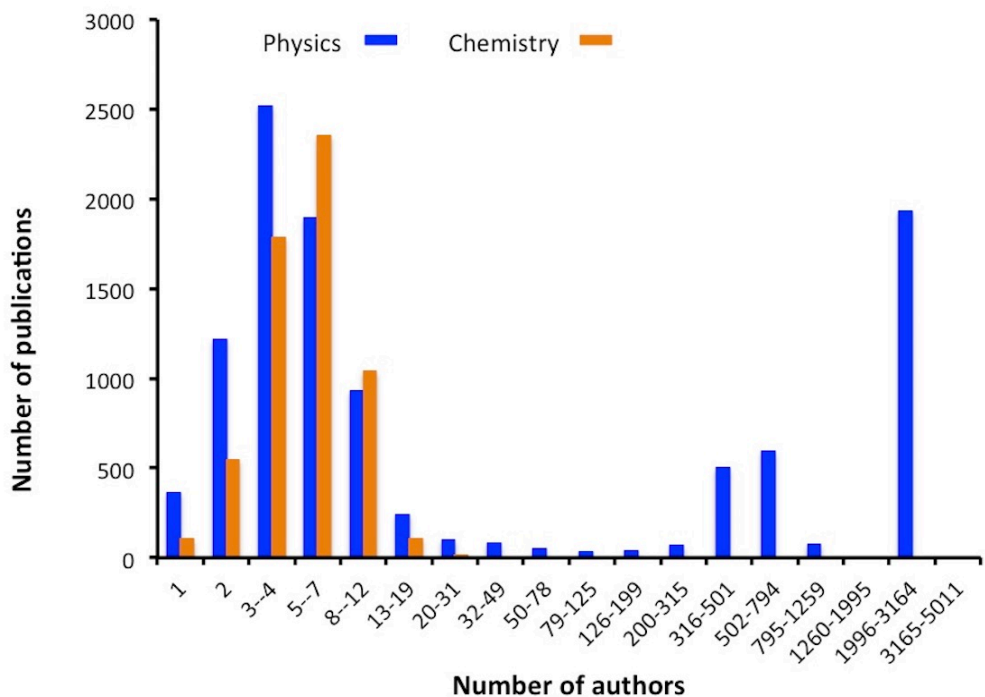

*Fig. 1. Distribution of the number of authors per publication in the WoS research areas of Physics and Chemistry in the UK universities recorded in Table 7. Publications in year 2012. Blue, Physics; Orange, Chemistry*

Table 7. Hyper-authored publications per university, percentage of publications in physics in 2012 exceeding 20, 50, and 100 authors

| University | Number or proportion of publications | | | |
| --- | --- | --- | --- | --- |
| | Total | > 20 (%) | > 50 (%) | > 100 (%) |
| University of Bath | 95 | 0.0 | 0.0 | 0.0 |
| University of Birmingham | 362 | 62.4 | 60.8 | 60.5 |
| University of Bristol | 472 | 34.1 | 33.7 | 32.8 |
| University of Cambridge | 1124 | 17.2 | 16.9 | 16.9 |
| University of Central Lancashire | 117 | 11.1 | 7.7 | 6.0 |
| University of Durham | 259 | 1.2 | 0.0 | 0.0 |
| University of Exeter | 81 | 0.0 | 0.0 | 0.0 |
| Imperial College London | 809 | 30.4 | 29.2 | 28.6 |
| University of Kent | 29 | 0.0 | 0.0 | 0.0 |
| King's College London | 155 | 1.3 | 0.6 | 0.6 |
| Lancaster University | 288 | 61.5 | 61.5 | 61.1 |
| University of Leeds | 183 | 3.3 | 3.3 | 2.7 |
| University of Leicester | 41 | 0.0 | 0.0 | 0.0 |
| University of Liverpool | 461 | 64.0 | 60.7 | 59.9 |
| University College London | 570 | 34.7 | 34.7 | 34.6 |
| Loughborough University | 99 | 0.0 | 0.0 | 0.0 |
| University of Manchester | 577 | 47.7 | 45.2 | 44.5 |
| University of Nottingham | 255 | 1.2 | 0.8 | 0.0 |
| University of Oxford | 990 | 28.3 | 26.9 | 26.0 |
| University of Portsmouth | 47 | 0.0 | 0.0 | 0.0 |
| Queen Mary University of London | 246 | 51.2 | 51.2 | 51.2 |
| Royal Holloway, University of London | 188 | 81.4 | 80.9 | 80.9 |
| University of Sheffield | 340 | 40.3 | 40.3 | 39.7 |
| University of Southampton | 369 | 26.0 | 25.5 | 25.5 |
| University of Surrey | 162 | 16.0 | 4.3 | 0.6 |
| University of Sussex | 198 | 65.2 | 65.2 | 65.2 |
| University of Warwick | 338 | 41.4 | 41.1 | 40.5 |
| University of York | 169 | 10.7 | 3.0 | 0.6 |
| University of Edinburgh + University of St Andrews | 602 | 40.7 | 37.9 | 36.9 |
| University of Glasgow | 442 | 62.4 | 59.7 | 58.8 |
| Heriot-Watt University | 129 | 0.0 | 0.0 | 0.0 |
| University of Strathclyde | 193 | 6.2 | 3.1 | 3.1 |
| Aberystwyth University | 14 | 0.0 | 0.0 | 0.0 |
| Cardiff University | 117 | 11.1 | 7.7 | 6.0 |
| Swansea University | 84 | 4.8 | 0.0 | 0.0 |
| Queen's University Belfast | 145 | 4.1 | 0.7 | 0.0 |

A possible explanation for the difficulties that were found in the calculation of the $e_p$ index in many universities could be that the hyper-author collaborations alter the double rank power law because normal and hyper-authored publications form two different populations regarding the distribution of citations. At global level, in 2012, the proportion of this type of collaboration was only 0.45% of all publications, which is insignificant and probably well integrated in the global lognormal citation distribution. In contrast, in 17 out of the 36 universities under study, the percentage of hyper-authored papers varied from 16.9% to 80.9% (Table 7) and the distortion of the lognormal citation distribution is possible.

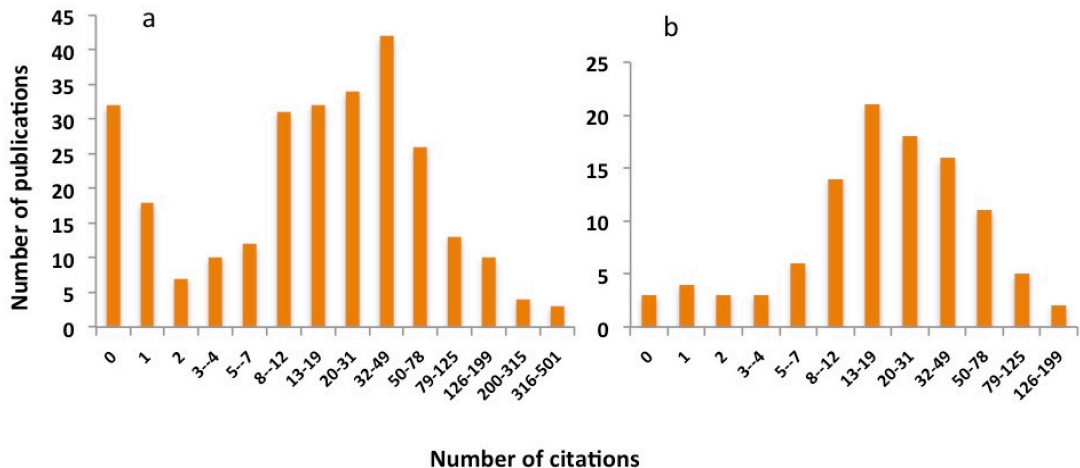

*Fig. 2. Distribution of citations to the papers published by the ATLAS and CMS collaborations (left) and LHCb and CDF collaborations (right)*

To test this possibility, we studied the citation distribution of the publications from the ATLAS and CMS collaborations (Fig. 2 left) and from the LHCb and CDF collaborations (Fig. 2 right). Omitting the lower tail with 0–2 citations, which clearly formed an independent population of the papers from the ATLAS and CMS collaborations, the rest of the two distributions resemble lognormal distributions with similar $\mu$ and $\sigma$ parameters: 3.2 and 1.1, and 3.1 and 0.9, respectively. In contrast, as a general fact, in universities without hyper-authored publications the $\mu$ parameter is smaller. For example, in the University of Durham, which has a fairly high $e_p$ index (Table 6), the $\mu$ and $\sigma$ parameters value 2.7 and 1.1, respectively, eliminating the 0–2 citation tail (distribution not shown). Obviously, the combination of two lognormal distributions with different parameters is not a lognormal distribution, which confirmed the aforementioned possibility of distribution distortion.

## 6. Discussion

### 6.1. Pros and cons of peer reviews and $e_p$ index-based indicators

The pros and cons of the use of bibliometric indicators or peer review for the research assessment of institutions have been extensively studied (Martin 2011; Wilsdon et al. 2015; Wouters et al. 2015) and the correlation of the $PP_{top10\%}$ indicator with the peer review of REF outputs has been demonstrated (Traag and Waltman 2019).

The aim of this study was to examine in more depth the percentile indicator that might eventually substitute for the peer review. If this substitution were made, university applications and sampling of publications would be unnecessary because the percentile indicator of the assessed university would be based on its whole production recorded in the WoS or other databases. The use of the whole production eliminates the sampling problem of the analysis of a low number of outputs, which may be insufficient to reveal the actual excellence of some universities (Section 3.1). When comparing the REF results with the percentile indicator, the effect of this problem—too low rating for high-level universities—is asymmetric because if it appears, it always increases the $PP_{top\,x\%}$/ PP(4*) ratios. The probability of appearance will be higher in fields with numerous groups and less strict panels. For example, if the PP(4*) indicator is equivalent to the top 1.0 percentile and research is performed by small groups, the existence of researchers with more than four 4-star-rated outputs in six years is unlikely. In contrast, if the PP(4*) indicator is equivalent to the top 10 percentile and research is performed by large groups, many researchers may publish more than four 4-star-rated outputs in six years.

In this study, the comparison of the results obtained by peer review and by using a percentile-based indicator is facilitated with the use of the $e_p$ index. The use of the percentile that exactly produces a ratio of 1.0 with peer assessments has the advantage of accurately indicating the level of the requirement that has been applied by the expert judges. Thus, different levels of peer requirements by different evaluating sub-panels across research areas are immediately revealed by the percentile that gets the closest ratio to 1.0 with peer assessment. Another minor but convenient advantage of the use of the percentile that exactly matches peer assessments is that proportional deviations from a ratio of 1.0 are more rapidly perceived than with other ratios. Furthermore, if the $PP_{top\ x\%}$/ PP(4*) ratio is 1.0, the $PP_{top\ x\%}$ is the indicator that allows an actual like-for-like substitution for peer assessment.

Another advantage of the use of the $e_p$ index is that it removes all problems of dichotomous distinction: for example, to decide whether to rate an output 4- or 3-star, which may be difficult. This problem does not apply to the $e_p$ index and the percentiles calculated from it, because the $e_p$ index is calculated from a fitting that implies a large number of publications. The dichotomous distinction problem will affect more sharply to universities with low number of outputs. If the number of 4- and 3-star-rated outputs is high, an internal compensation of opposite mistakes can be expected.

The results obtained in three research areas allow the pros and cons of the method to be determined more clearly.

**6.1.1. Chemistry**

In the UOA of Chemistry, using the 2.8 percentile, in 21 out of 26 universities, the $PP_{top\ 2.8\%}/ PP(4*)$ ratio deviates moderately from 1.0, from 1.67 to 0.66; in 15 universities the ratio varied from 1.10 and 0.82 (Table 2). Considering all universities, there is an upper tail of five universities in which peer evaluations resulted much less favourably than percentile evaluation (ratios > 1.67). This asymmetric upper tail may be due to the sampling method in REF (Sections 3.1 and 5.1) and could explain the cases of the Imperial College of London and University of Cambridge and eventually of the University of Bath. The presence in this tail of Newcastle University must have another reason, perhaps the dichotomous distinction problem described above. In fact, in Newcastle University four outputs were rated 4-star and 72 were rated 3-star. Thus, if a few real 4-star outputs were rated 3-star the $PP_{top\ 2.8\%}/ PP(4*)$ ratio would increase above the 1.0 value. Fortunately, this type of failure would be easily detected if is studied by expert reviewers.

The range of deviations from 1.10 to 0.82, which applies to 15 universities, could be taken as the normal variability that is inherent to any type of evaluation. In fact, performing ratios increases the variability of the original data; in a ratio range from 1.10 to 0.82, the expected variability of the two indicators is low. For example, with a variability of ±10% in both types of data, the expected variability of the ratios would be higher than that from 1.10 to 0.82. Most likely, most experts in research assessment would consider that a variability of ±10% in research assessments is positively surprising.

### 6.1.2. Economics and business

In the two UOAs # 18 and # 19, we could only test 16 universities against a percentile indicator because in many of the universities in REF the number of publications was too low to allow a robust calculation of the $e_p$ index. However, the study of these 16 universities provided informative results.

In these universities, the 9.0 percentile was the most convenient to compare to the REF results. The variability of the $PP_{top\ 9.0\%}/PP(4*)$ ratio was rather high (mean = 1.03, standard deviation = 0.43), but in contrast with Chemistry, the ratios were perfectly distributed around the mean (Table 4). Furthermore, if we consider the ratios in the two universities with the higher number of Nobel laureates in Economic Sciences—University of Cambridge and London School of Economics and Political Science, 1.21 and 0.70, respectively—the same number of universities exhibited ratios higher than 1.21 and ratios lower than 0.70. This distribution suggests that divergences between the peer and percentile assessments are not the result of any specific bias.

Several causes can explain these divergences. Two of them might be that 2.6% of the submitted outputs are not journal articles (Wilsdon et al. 2015, p. 154) and that the WoS list of journals in the research areas of Business & Economics and Operations Research & Management Science may not cover all the journals where the submitted outputs are published. Although the proportion of these uncovered publications seems too low to explain the variability observed, it is possible that it affects some universities more than others, as they distribute unevenly across universities.

To go further in the analysis of the divergences between peer review and the percentile indicator would require a specific analysis by experts in economics and business who study the 4-star-rated outputs and their number of citations. Aside from this issue, some observations suggest that in the field of economics and business the selection of the outputs submitted and the evaluation of these outputs may be different to those in chemistry or physics. In the first place, the $PP_{top\ x\%}$ that is equivalent to the $PP(4*)$ indicator is higher in the UOAs of Economics and Business versus Chemistry or Physics, 9% versus 2.8% and 1.1%, respectively. Furthermore, and probably related, the comparison of the $PP(4*)$ columns in Tables 2, 4, and 6 show notable differences between the research areas, because it is evident that the proportion of 4-star-rated outputs versus the total number of articles recorded in the WoS is notably higher in Economics and Business than in Chemistry and Physics (the means are 11.7, 2.7, and 2.3, respectively).

Although the deviations of the $PP_{top\ 9\%}/PP(4*)$ ratio from 1.0, ranging from 1.85 to 0.40 seem large, they are not so large; they could be expected from a variation of the data of ±35%. However, at the level of this study it is not possible to conclude whether the PP$_{top\ 9.0\%}$ indicator is a reasonably like-for-like substitute for the PP(4*) indicator and a straightforward substitution is doubtful. However, it seems likely that a study by experts in economics and business could reach a positive conclusion perhaps suggesting simple complements for the bibliometric analysis.

### 6.1.3. Physics

In the UOA of Physics we could only study 12 out of 40 universities. In part, this is because in some universities the number of WoS articles (Table 5) was insufficient for a robust calculation of the $e_p$ index, but also because in many cases the percentile-based double rank distribution could not be fitted to a power law. The cause of this impossibility is the notable proportion of hyper-authored papers (Fig. 1) that occurs in many universities (Table 7), which is due to wide participation in international collaborations.

The study of the ATLAS and CMS collaborations, which involve around 3,000 authors per paper, and the LHCb and CDF collaborations, which involve 300–800 authors per paper, showed that their citation distributions could be fitted to lognormal distributions, but with $\mu$ parameters that are higher to those of papers with a low number of authors—the question of whether this difference is due to the high number of authors or to the scientific characteristics of subject is out of the scope of this study. In the global distribution of citations, the proportion of hyper-authored papers (all types of them) is very low (0.45% in 2012), which strongly suggests that they do not have a significant influence in the global lognormal citation distribution in the WoS research area of Physics. In contrast, in many UK universities the proportion is much higher (20–80%; Table 7), which distorts the lognormal distribution of citation and subsequently the percentile-based double rank distribution could not be fitted to a power law.

The conclusion that can be drawn from these results is that the bibliometric evaluation of these papers must be done independently from the evaluation of the other papers with a low number of authors because they belong to two independent citation universes. Furthermore, a certain agreement about how to perform the combination of the evaluations of both types of papers must be reached because the proportion of normal and hyper-authored papers varies across universities (Table 7) and many of these hyper-authored papers are listed in several universities.

It is worth noting that the aforementioned difficulties are not exclusive to bibliometric evaluations, they also apply to peer evaluations. An example of two publications submitted as outputs in REF by the same university illustrates the issue. The first publication describes an efficient solar cell (Liu et al. 2013) and is authored by three researchers, two of whom are staff members of the university. Most peers will rate this publication as 4-star and it can be attributed to only one university. The second publication is an ATLAS Collaboration research about the Higgs boson (Aad et al. 2013), which is authored by 2,922 researchers who belong to 179 institutions, including 13 UK universities; nine of its authors are staff members of the university under consideration and the publication is an output that was listed twice in the implied university. Many particle physicists would probably rate this publication as a 4-star, which could be done in 13 universities, and more than once by the same university. For evaluative purposes, these two publications are so different that it seems that an equitable judgment of both in a comparative way is an almost impossible task, unless that, as mentioned above, a method of evaluation has been previously agreed.

The discrepancies between peer and bibliometric evaluations in the universities of Nottingham and York, $PP_{top\ 1.1\%}/PP(4^*)$ ratios of 0.33 and 0.49 (Table 6), seem high although it cannot be ruled out that it is normal variability. There is nothing in these universities that could explain the notable deviation from the 1.0 ratio: (i) the number of publications is sufficient for reliable fittings; (ii) the numbers of submitted outputs is high, 137 and 193, respectively; (iii) the values of the $e_p$ index are normal, $\approx 0.1$, and (iv) the number of multi-authored publications is very small. Therefore, the high rating of these universities by the experts with reference to the percentile indicator (ratios much lower than 1.0) deserves a specific analysis of the two indicators by experts in physics, which should clarify the deviations.

## 6.2. Top percentile equivalence of 4-star peer ratings reveals international relevance

The use of the $e_p$ index in this study has allowed the characterization of the 4-star or world-leading quality in an international context. In the case of Chemistry this top quality is equivalent to the top 2.8% of cited papers, in the case of Economics and Business the equivalence is to the top 9.0% of cited papers, and in the case of Physics, the equivalence is with the top 1.1 % of cited papers.

These important differences between UOAs suggest that experts in different research areas keep different criteria regarding the concept or *world leading research* (4-star rating). The use of the $e_p$ index could serve to homogenize evaluations across these UOAs. This homogenization might not be strictly necessary, but it seems convenient to have common criteria for universities that are specialized in different research areas.

## 6.3. Probability of publishing a very highly cited paper

An additional and main advantage of evaluations with the $e_p$ index is that it allows the immediate calculations of the probabilities of achieving the publication of highly cited papers located in the 0.01 or any other percentile. The convenience of the calculation of the probability or expected frequency at these low percentile seems reasonable if the evaluation tries to determine the capacity of the system to achieve important breakthroughs (Rodríguez-Navarro and Brito 2019); Tables 2, 4, and 6 report the value of the $PP_{top\ 0.01\%}$ indicator in many universities. Although comparatively this indicator does not change the judgements that can be made with the $e_p$ index, because it equals the value of this index to the power of four (Eq. 1), it has the advantage of providing the actual figures for achieving breakthroughs at a concrete level. These figures might eventually serve to discuss funding differences between universities. If random papers in chemistry in the Universities of Cambridge and York have probabilities of around 0.1 to reach the top 0.01 citation percentile and random papers in some other universities are 10 or even 100 times lower, university administrators might like to take into account these differences.

With regard to the $e_p$ index values, in the three research areas here studied, the UK universities lag behind the MIT and Princeton University (Tables 2, 4, and 6). Although at a first glance, it seems that physics is ahead of chemistry, and economics and business, it is more probable that the three cases are similar, with top UK universities exhibiting $e_p$ index values of around 0.18–0.20 while world-leading universities exhibit index values of 0.22–0.25 $e_p$ index values.

Although this comparison seems informative, it must be interpreted with caution because international comparisons of universities based on the $e_p$ index are complex (Rodríguez-Navarro and Brito 2019). In addition to differences due to differences in research policy, the MIT and Princeton University are exceptional research universities that exist in a big country, the USA, where there are many top research universities such as, for example, Cornell University, University of Wisconsin at Madison, University of Illinois at Urbana-Champaign, and many others. This circumstance and the high mobility of researchers make possible the existence of the MIT and Princeton University and a few others with an exceptionally high $e_p$ index. The UK is smaller than the USA and it is probably impossible that its top universities can achieve $e_p$ index values similar to those of the MIT and Princeton University. However, this impossibility for countries smaller than the USA to have very high $e_p$ index universities does not imply that these countries cannot be very competent in research (Rodríguez-Navarro and Brito 2019).

### 6.4. Non-studied research areas

In this study only three research areas have been included. The intention is that these three areas reveal the framework, and the advantages and limits of a like-for-like substitution of a bibliometric indicator for peer review, laying the groundwork for more extensive studies.

The findings in the field of chemistry suggest that this field is a good candidate to be evaluated with a percentile indicator. However, it is pending a study by experts of the discrepancy observed between peer and bibliometric evaluations in the University of Newcastle. It can be expected that this study will reveal a specific problem rather than a general one.

According to previous experience (Brito and Rodríguez-Navarro 2018b; Rodríguez-Navarro and Brito 2018) the field of chemistry studied here is probably representative of many fields in natural and formal sciences, and technological fields for assessment with percentile indicators. This conclusion also applies to the papers in physics with a low number of authors (Fig. 1), with the pending study of the universities of York and Nottingham. The evaluation of multi-authored papers needs further studies and perhaps agreements. The study of the evaluation of multi-authored papers in physics might also serve as a model for other multi-authored papers in clinical medicine and perhaps in other areas.

The fields of economics and business (UOAs of Economics and Econometrics and Business and Management Studies in REF) might represent a limit in the substitution of a bibliometric indicator for peer review in social sciences. Although the variability of the $PP_{top\ 9\%}/PP(4*)$ ratio is still compatible with the general difficulties of performing a research evaluation, it might also respond to specific difficulties whose existence needs to be ruled out. Other fields in social sciences or humanities will require specific studies.

An additional issue is that in order for a bibliomeric indicator to reliably substitute peer review extensively, the number of research fields—equivalent to current OUAs—must be regrouped in order that all universities publish a sufficient number of papers to obtain the indicator robustly. Alternatively, a statistical approach that allows the study of several years together might solve this problem.

**7. Conclusions**

The $e_p$ index and percentile indicators calculated from it provide a solid basis for the selection of a bibliometric indicator that may substitute for the peer review of publications in future UK REF and in research assessments in other countries. However, several steps must be performed before the substitution can be applied successfully. These steps include (i) deciding the research areas to which the $e_p$ index approach can be applied; (ii) in these areas, finding explanations for the specific discrepancies that are found in REF between bibliometric and peer review evaluations; and (iii) deciding the grouping of research fields in order that the $e_p$ index can be robustly obtained.

These studies might appear laborious in absolute terms but not so much considering the context, because it could be expected that the substitution of a bibliometric indicator for a peer review process as meticulously elaborated as the REF could not be achieved straightforwardly. The advantage is that these studies can be performed using the REF results, where the maximum effort has already been performed. Furthermore, the benefits of substituting an $e_p$ index-based indicator for the complex and onerous process of peer review might remove the possibility of giving up performing the evaluations of research institutions (Martin 2011), which applies not only to the UK but also to many other countries. It is worth noting that the risk of changing the research system in the process of measuring it (Martin 2011) is small in evaluations with the $e_p$ index (Rodríguez-Navarro and Brito 2019).

## Acknowledgment

This work was supported by the Spanish Ministerio de Economía y Competitividad, grant number FIS2017-83709-R.

## References

Aad, G, et al. (2013), 'Evidence for the spin-0 nature of the Higgs boson using ATLAS data', *Physics Letters B*, 726, 120-44.
Acheson, R M (1973), 'Blood pressure in a national sample of U.S. adults: Percentile distribution by age, sex and race', *International Journal of Epidemiology*, 2, 293-301.

Albarrán, P, Ortuño, I, and Ruiz-Castillo, J (2011), 'Average-based versus high- and low-impact indicators for the evaluation of scientific distributions', *Research Evaluation,* 20, 325-39.

Birnholtz, J P (2006), 'What does it means to be an author? The intersection of credit, contribution, and collaboration in science', *Journal of the American Society for information Science and Technology,* 57, 1758-70.

Bonaccorsi, A (2007), 'Explaining poor performance of European science: institutions versus policies', *Science and Public Policy,* 34, 303-16.

Bornmann, L and Marx, W (2015), 'Methods for the generation of normalized citation impact scores in bibliometrics: Which method best reflects the judgements of experts?', *Journal of Informetrics,* 9, 408-18.

Bornmann, L, Leydesdorff, L, and Mutz, R (2013a), 'The use of percentile rank classes in the analysis of bibliometric data: opportunities and limits', *Journal of Informetrics,* 7, 158-65.

Bornmann, L, Leydesdorff, L, and Wang, J (2013b), 'Which percentile-based appoach should be preferred for calculating normalized citation impact values? An empirical comparison of five approaches including a newly developed citation-rank approach (P100)', *Journal of Informetrics,* 7, 933-44.

Bornmann, L, Ye, A, and Ye, F (2018), 'Identifying landmark publications in the long run using field-normalized citation data', *Journal of Documentation,* 74, 278-88.

Brito, R and Rodríguez-Navarro, A (2018a), 'Research assessment by percentile-based double rank analysis', *Journal of Informetrics,* 12, 315-29.

--- (2018b), 'The USA is an indisputable world leader in medical and biotechnological reserach', *Preprint at arXiv,* :1807.01225.

--- (2019), 'Evaluating reserach and researchers by the journal impact factor: Is it better than coin flipping?', *Journal of Informetrics,* 13, 314-24.

Cronin, B (2001), 'Hyperauthorship: A postmodern perversion or evidence of a structural shift in scholarly communication practises?', *Journal of the American Society for information Science and Technology,* 52, 558-69.

De-Bellis, N (2009), *Bibliometrics and Citation Analysis - From the Science Citation Index to Cybermetrics* (The Scarecrop Press, Inc. Lanham, Maryland).

Dosi, G, Llerena, P, and Labini, M S (2006), 'The relationships between science, technologies and their industrial exploitation: An illustration through the myths and realities of the so-called 'European Paradox'', *Research Policy,* 35, 1450-64.

Dunaiski, M, Visser, W, and Geldenhuys, J (2016), 'Evaluating paper and authors ranking algorithms using impact and contribution awards', *Journal of Informetrics,* 10, 392-407.

Gallman, R E (1969), 'Trends in the size distribution of welth in the nineteenth century: Some speculations', in L Soltow (ed.), *Six Papers on the Size Distribution of Wealth and Income* (Cambridge, Massachusetts: National Bureau of Economic Research).

Godin, B (2006), 'On the origins of bibliometrics', *Scientometrics,* 68, 109-33.

Harnad, S (2008), 'Validating research performance metrics against peer rankings', *Ethics in Science and Environmental Politics,* 8, 103-07.

--- (2009), 'Open access scientometrics and the UK research assessment exercise', *Scientometrics,* 79, 147-56.

Herranz, N and Ruiz-Castillo, J (2013), 'The end of the "European Paradox"', *Scientometrics,* 95, 453-64.

King, C (2012), 'Multiauthor papers: Onward and upward', *Science Watch Newsletter*. <http://archive.sciencewatch.com/newsletter/2012/201207/multiauthor_papers/>.

Kreiman, G and Maunshell, J H R (2011), 'Nine criteri for a measure of scientific output', *Frontiers in Computational Neuroscience,* 5, Article 48.

Kuhn, T (1970), *The structure of scientific revolutions* (Chicago: University of Chicago Press).

Liu, M, Johnston, M B, and Snaith, H J (2013), 'Efficient planar heterojunction perovskite solar cells by vapour deposition', *Nature,* 501, 395-98.

Martin, B R (2011), 'The Research Excellence Framework and the 'impact agenda': are we creating a Frankenstein monster?', *Research Evaluation,* 20, 247-54.

Mingers, J and Leydesdorff, L (2015), 'A review of theory and practice in scientometrics', *European Journal of Operational Research,* 246, 1-19.

Persson, O, Glänzel, W, and Danell, R (2004), 'Inflationary bibliometric values: The role of scientific collaboration and the need for relative indicators in evaluative studies', *Scientometrics,* 60, 421-32.

Press, W H, Flannery, B P, Teulosky, S A, Vetterling, W T (1989), '*Numerical Recpies, Fortran Version*' (Cambridge University Press, Cambridge, United Kingdom)

REF2014 (2011), 'Assessmentframework and guidance on submissions', *REF 02.2011*.

Régibeau, P and Rockett, K E (2016), 'Research assessment and recognized excellence: simple bibliometrics for more efficient academic research evaluations', *Economic Policy,* 31, 611-52.

Rodriguez-Navarro, A and Narin, F (2018), 'European paradox or delusion-Are European science and economy outdated?', *Science and Public Policy,* 45, 14-23.

Rodríguez-Navarro, A (2012), 'Counting highly cited papers for university research assessment: conceptual and technical issues', *PLoS One,* 7(10), e47210.

Rodríguez-Navarro, A and Brito, R (2018), 'Technological research in the EU is less efficient than the world average. EU research policy risks Europeans' future', *Journal of Informetrics,* 12, 718-31.

--- (2019), 'Probability and expected frequency of breakthroughs – basis and use of a robust method of research assessment', *Scientometrics,* (Early view DOI https://doi.org/10.1007/s11192-019-03022-1).

Rossi, P, Strumia, A, and Torre, R (2019), 'Bibliometric for collaboration works', *Preprint at arXiv:1092.01693v1*.

Salter, A J and Martin, B R (2001), 'The economic benefits of publicly funded basic research: a critical review', *Research Policy,* 30, 509-32.

Sonnenwald, D H (2007), 'Scientific collaboration', *Annual Review of Information Science and Technology,* 41, 643-81.

Tijssen, R J W, Visser, M S, and van-Leeuwen, T N (2002), 'Benchmarking international scientific excellence: Are highly cited research papers an appropriate frame of reference?', *Scientometrics,* 54, 381-97.

Traag, V A and Waltman, L (2019), 'Systematic analysis of agreement between metrics and peer review in the UK REF', *Palgrave Communications,* 5, 29.

Waltman, L (2016), 'A review of the literature on citation impact indicators', *Journal of Informetrics*, 10, 365-91.

Waltman, L and Schreiber, M (2013), 'On the calculation of percentile-based bibliometric indicators', *Journal of the American Society for information Science and Technology*, 64, 372-79.

Waltman, L and van-Eck, N J (2013), 'A systematic empirical comparison of different approaches for normalizing citation impact indicators', *Journal of Informetrics*, 7, 833-49.

--- (2015), 'Field-normalized citation impact indicators and the choice of an appropriate counting mehod', *Journal of Informetrics*, 9, 872-94.

Weinberg, A M (1962), 'Criteria for scientific choice', *Minerva*, 1, 158-71.

--- (1964), 'Criteria for Scientific choice II: the two cultures', *Minerva*, 3, 3-14.

Wilsdon, J, et al. (2015), *The metric tide: Report of the independent review of the role of metrics in research assessment and management*.

Wouters, P, et al. (2015), *The Metric Tide: Literature Review (Supplementary Reprot I to the Independent Review of the Role of Metrics in Research Assessments and Management)* (HEFCE DOI: 10.13140/RG.2.1.5066.3520).

Wuchty, S, Jones, B F, and Uzzi, B (2007), 'The increasing dominance of teams in production of knowledge', *Science*, 316, 1036-39.